\documentclass[prl,nofootinbib,noshowkeys,twocolumn,superscriptaddress]{revtex4}
\usepackage[latin1]{inputenc}
\usepackage{pstricks,pst-node,pst-text,pst-3d,pslatex}
\usepackage[T1]{fontenc}
\usepackage{amsmath}

\usepackage{graphicx,amsfonts}
\usepackage{amsmath}

\usepackage[english]{babel}
\newcommand{\be}{\begin{eqnarray}}
\newcommand{\ee}{\end{eqnarray}}
\newcommand{\la}{\langle}
\newcommand{\ra}{\rangle}

\begin{document}

\title{Accelerated Sampling of Boltzmann distributions}

\author{Henri Orland}

\email{henri.orland@cea.fr}

\homepage{http://ipht.cea.fr}

\affiliation{Institut de Physique Théorique\\
CEA, IPhT and CNRS, URA 2306\\
F-91191 Gif-sur-Yvette, France}

\begin{abstract}
The sampling of Boltzmann distributions by stochastic Markov processes,
can be strongly limited by the crossing time of high (free) energy barriers. As a
result, the system may stay trapped in metastable states, and the
relaxation time to the equilibrium Boltzmann distribution may be very
large compared to the available computational time. In this paper,
we show how, by a simple modification of the Hamiltonian, one can
dramatically decrease the relaxation time of the system, while retaining
the same equilibrium distribution. 
The method is illustrated on the case of the one-dimensional double-well potential.
\end{abstract}
\maketitle
The sampling of complex and rugged landscapes is a problem of major importance in many different fields of Science, such as physics, biology, optimization,
probability theory, etc \cite{optimization}. In statistical physics, simulation
methods rely on an efficient sampling of the important regions of phase
space.  The quality of the sampling guarantees a reliable computation
of observables. 

There are many Markov processes which can in principle sample  Boltzmann
distributions. Among the most used are Monte Carlo and  Molecular dynamics methods, the Langevin equation \cite{MC}, etc.  combined with simulated annealing \cite{SA}  or  quantum annealing \cite{QA}. These processes mimick the dynamical sampling of the phase space, and in many cases, the microscopic time (time step) must be chosen very small compared to the relaxation time in order for the system to indeed sample the Boltzmann distribution and reach its thermodynamic equilibrium. A typical example of this difficulty is the famous "Protein Folding" problem \cite{protein}. In this case, given a microscopic Hamiltonian for the protein, one tries to find the room temperature conformation of the molecule by running one of the above mentioned algorithms. However, the microscopic time  (discretization time) is typically of the order of $10^{-15}$s while the typical folding (relaxation) time runs from milliseconds to seconds. The longest simulations available as of now for very short proteins  run in the $10^{-7}$s range, still very far from the equilibration time. The same difficulty occurs in many disordered systems (spin-glasses), in amorphous systems, glasses, polymers, etc.
It is thus very important to find a way to dramatically accelerate the sampling, so as to make it efficient with present days computers.

In this paper, we assume that the system is subject to an overdamped Langevin dynamics (also called Brownian Dynamics). We show how by simply modifying the associated 
Schr\"odinger equation, one can reduce very substantially the relaxation time to equilibrium.

For the sake of simplicity, we will specialize to the case of a one-dimensional particle in a potential well $U(x)$. All the following can be trivially extended to the case of any number of particles in any dimensions.

Consider a particle subject to the overdamped Langevin dynamics in a potential $U(x)$
\be
{\dot x} = - \frac {D}{k_BT} \frac{\partial U}{\partial x} + \eta (t)
\ee
where $D$ is the diffusion constant of the particle, $k_B$ is the Boltzmann constant, $T$ is the temperature, and $\eta(t)$ is a white Gaussian noise such that 
\be
\la \eta(t) \ra &=& 0\\
\la \eta(t) \eta(t') \ra &=& 2 D  \delta (t-t')
\ee

 This Langevin equation describes a Markov process, and its probability distribution function (pdf) $P(x,t)$ satisfies the Fokker-Planck equation
 \be
 \frac {\partial P}{\partial t} = D \frac{\partial}{\partial x}\left ( \frac{\partial P}{\partial x} + \frac{\partial (\beta U)}{\partial x} P \right )  
 \ee
where $\beta  = 1/ k_BT$. The function defined by 
\be
\Psi(x,t) = e^{\beta U(x)/2} P(x,t)
\ee
satisfies a Schr\"odinger equation \cite{schrodinger}
\be
\label{schrod}
\frac{\partial \Psi}{\partial t} = -H \Psi(x,t)
\ee
with
\be
\label{ham}
H = -D \frac{\partial^2}{\partial x^2} + D V_e (x)
\ee
and
\be
\label{veff}
V_e(x) = \left(\frac{\beta}{2}\frac{ \partial U}{\partial x}\right)^2 - \frac{\beta}{2} \frac{\partial^2 U}{\partial x^2}
\ee

The solution of eq. (\ref{schrod}) can be formally expanded as
\be
\Psi(x,t) = \sum_{\alpha \in N} e^{-E_{\alpha}t} \Psi_{\alpha}(x)  c_{\alpha}
\ee
where $\Psi_{\alpha}$ and  $E_{\alpha}$ are the normalized eigenstates and eigenvalues of $H$ 
\be
H | \Psi_{\alpha}\ra = E_{\alpha} | \Psi_{\alpha} \ra
\ee
and the $c_{\alpha}$ represent the initial condition on the wavefunction $\Psi(x,0)$
\be
c_{\alpha} = \int dx \Psi_{\alpha}(x) \Psi(x,0)
\ee
Note that since $H$ is a Hermitian operator, the eigenstates $\Psi_{\alpha}$ are orthogonal with each other.

It is well known that the ground state of the Hamiltonian (\ref{ham}) is given by
\be
\label{psi}
\Psi_0(x) = \frac {e^{-\beta U(x)/2}}{\sqrt{Z_0}}
\ee
where $Z_0$ is the partition function of the original system
\be
Z_0 = \int dx e^{-\beta U(x)}
\ee
Indeed, it is easily checked that
\be
H |\Psi_0 \ra =0
\ee
which means that $\Psi_0$ is an eigenstate of $H$ with zero eigenvalue.
In addition, since $\Psi_0 (x)$ is strictly positive, it is necessarily the ground state of $H$, and thus all other eigenvalues are positive $E_{\alpha} > 0$.

We can thus expand $\Psi(x,t)$ on the basis of eigenstates as
\be
\Psi(x,t) = e^{-\beta U(x)/2} c_0 + e^{-E_1 t} \Psi_1(x) c_1 + \dots
\ee
and we see that the relaxation time to the Boltzman distribution is given by
\be
\tau_R = \frac{1}{E_1}
\ee
where $E_1$ is the smallest non zero eigenvalue. Note that the smallest the eigenvalue, the largest the relaxation time.

It is well-known that if there are high energy barriers, the gap between the lowest energy states is exponentially small in the barrier height. For instance, in the double-well potential, the energy splitting between the two lowest eigenvalues is known to be exponentially small in the barrier height. We see therefore that large energy barriers in phase space imply a very small energy gap and thus very long relaxation times. Thus any dynamics (Langevin, Monte Carlo, Molecular Dynamics) that mimicks the real dynamics of the system will be subject to very long relaxation times and very slow relaxation rates.

In order to cure the problem of the small gap, we transform the Hamiltonian $H$ into a new Hamiltonian $H_{\lambda}$ which has exactly the same eigenfunctions as $H$ but a large gap. We are able to do that thanks to the fact that we know exactly the ground state $\Psi_0$ of the Hamiltonian.

Consider the Hamiltonian operator
\be
\label{hammod}
H_{\lambda} = H - \lambda P_0
\ee
where $P_0$ is the projector onto  the ground state
\be
 P_0&=& |\Psi_0\ra \la \Psi_0| \\
 P_0(x,y)&=& \frac {e^{-\frac{\beta}{2}(U(x)+U(y))}}{Z_0} \label{projector}
 \ee
and $\lambda >0 $ is an arbitrary constant. The eigenstates and eigenvalues of $H_{\lambda}$ are the same as those of $H$, except for the ground state energy which is shifted by $\lambda$. 
Indeed, we have
\be
\label{spectrum1}
H_{\lambda} |\Psi_0\ra &=& - \lambda |\Psi_0\ra
\\
\label{spectrum2}
H_{\lambda} |\Psi_{\alpha} \ra &=& E_{\alpha}  |\Psi_{\alpha}\ra \ \ \  {\rm for} \ \ \ \alpha \ge 1
\ee
and thus the new energy gap, which determines the relaxation time and rate is given by
\be
\label{delta}
\Delta = E_1 + \lambda 
\ee
and can be made very large by increasing $\lambda$.

The price to pay for increasing the gap is that the new Hamiltonian $H_{\lambda}$ is no more local. Also, it cannot be derived from a Langevin equation through the process described above. However, as we now show, it can be easily sampled, due to its very simple structure.

In order to sample $\Psi_0$, one may use quantum Monte Carlo methods. In the following, we use the Feynman path integral representation. Consider the wavefunction
\be
\label{phi}
\Phi(x,t) = e^{-\lambda t} \ e^{-H_{\lambda} t} \Phi_0 (x)
\ee
where $\Phi_0$ is the initial pdf.
At large time, $t >> 1/\Delta$, we have
\be
\Phi(x,t) \sim e^{-\beta U(x)/2} c_0 + e^{-\Delta t} \Psi_1(x) c_1 + \dots
\ee
and thus the system relaxes to a Boltzmann distribution at temperature $ 2 T$  with a smaller relaxation time given by
\be
\tau_{\lambda} = \frac{1}{E_1 + \lambda}
\ee

Using a Trotter-like approach, we discretize the exponential evolution operator in (\ref{phi}) 
\be
\la x| e^{-  \epsilon H_{\lambda}} | x'\ra = \la x| e^{-\epsilon H } \ e ^{+ \epsilon \lambda P_0}|x'\ra
\ee
where $\epsilon$ is a small timestep and the exponential factorizes exactly because $H$ commutes with the projector $P_0$. Since $P_0$ is a projector, we have
\be
e ^{+ \epsilon \lambda P_0}= 1 + (e^{\epsilon \lambda}-1) P_0
\ee
and thus
\be
\label{trotter}
\la x| e^{-  \epsilon H_{\lambda}} | x'\ra = \la x| e^{-\epsilon H } |x'\ra + (e^{\epsilon \lambda}-1) \Psi_0(x) \Psi_0(x')
\ee

Using the standard expression for the kernel of the infinitesimal evolution operator (valid to third order in $\epsilon$), we have
\be
\label{cond}
\la x| e^{-\epsilon \lambda -  \epsilon H_{\lambda}} | x'\ra &\simeq& 
\frac{ e^{- \epsilon \lambda}}{\sqrt{4 \pi D \epsilon}} \ e^{- \frac {(x-x')^2}{4 D \epsilon} - \frac{\epsilon}{2} (V_e(x)+V_e(x'))}  \nonumber \\
 &+& \frac{ (1 - e^{-\epsilon \lambda})}{Z_0} e^{-\beta (U(x)+U(x'))/2}
\ee

All the terms in the above equation are known explicitly and thus can be evaluated numerically, except for the partition function $Z_0$.
This transfer matrix represents the conditional probability for the particle to be at point $x'$ at time $t+\epsilon$ given that it was at point $x$ at time $t$.
Note that there are two types of possible moves: i) either the particle goes from $x$ to $x'$ at a distance of order $\sqrt{\epsilon}$ as in usual Monte Carlo, in which case both terms in (\ref{cond}) may contribute, or ii) the particle moves to $x'$ at a large distance from $x$, in which case the first term is negligible and only the second may contribute. This second term thus introduces some non-locality in  the stochastic sampling process, and allows for large moves which satisfy detailed balance. 

The value of $Z_0$ is a priori not known. 
However, the precise value of $Z_0$ is in fact irrelevant. One obvious way to see it is through eq.(\ref{hammod}) and (\ref{projector}) which clearly show that the value of $Z_0$ can be absorbed in the definition of $\lambda$. A more explicit way to deal with $Z_0$ is the following:
assume we divide $Z_0$ by an arbitrary constant $A$ in (\ref{cond}) and consider the modified infinitesimal evolution operator
\be
M =e^{- \epsilon \lambda} e^{-\epsilon H }+ (1- e^{-\epsilon \lambda})\ A\  Z_0 \ \Psi_0(x) \Psi_0(x')
\ee
with matrix element
\be
\label{condmod}
M(x,x') &=&
\frac{ e^{- \epsilon \lambda}}{\sqrt{4 \pi D \epsilon}} \ e^{- \frac {(x-x')^2}{4 D \epsilon} - \frac{\epsilon}{2} (V_e(x)+V_e(x'))}  \nonumber \\
 &+& (1 - e^{-\epsilon \lambda})\ A\  e^{-\beta (U(x)+U(x'))/2}
\ee
where $A$ is an arbitrary constant.

Again, $|\Psi_0 \ra $ is the maximal eigenstate of $M$ with eigenvalue $\lambda_0 = e^{-\epsilon \lambda} + (1 - e^{-\epsilon \lambda}) A Z_0$. 
A standard method for sampling the evolution operator (\ref{condmod}) is the diffusion Monte Carlo method  \cite{MC}. In this method, an initial population of points, representative of the initial guess for the probability distribution, is replicated or deleted according to (\ref{condmod}). The iteration of the process guarantees that the population of points will asymptotically be distributed according to the ground state wave function $|\Psi_0 \ra $. The fact that the largest eigenvalue of $M$ is not $1$ is corrected by properly rescaling the population size to keep it approximately constant.

The rate of convergence is given by the ratio $\lambda_1 / \lambda_0$ of the first eigenvalue to the maximal eigenvalue
\be
\label{convrate}
\frac{\lambda_1}{\lambda_0} = \frac{e^{-\epsilon E_1}}{1+(e^{\epsilon \lambda} - 1) A Z_0}
\ee
The smaller this ratio, the faster the convergence to the Boltzmann distribution.

As far as the choice of $\lambda$ is concerned, note that all the equations are valid for any value of $\lambda$ and thus one might be tempted to use a very large value of $\lambda$ to accelerate convergence. However, this would favour sampling of distant minima, at the expense of sampling each minimum locally. It is difficult to assess what is the optimal value of $\lambda$ for an efficient local and non-local sampling. A small value of $\lambda$ would allow the sampling of only local minima, whereas a large value of $\lambda$ would sample only distant minima. Using the same strategy as in continuous Monte Carlo methods (for fluids for instance) \cite{MC}, it seems reasonable to adjust the value of  $\lambda$ so that typically half of the moves are local and half non-local.

To illustrate the method, we have tested it on the one-dimensional double-well potential
\be
U(x)=(x^2-1)^2
\ee

In that case, the effective potential $V_{e}$ is a polynomial of degree 6, and has 3 minima at low temperature.

First, it is easy to check numerically that the dependence of the spectrum of the transfer matrix (\ref{cond}) indeed satisfies equations (\ref{spectrum1},\ref{spectrum2}) and that the gap follows eq. (\ref{delta}).

 We have performed a diffusion Monte Carlo calculation \cite{MC}. We have fixed the constant $A=1$ in eq.(\ref{condmod}). We start with a population of 20000 points localized in one minimum, say $x=-1$. For each $x$, we draw the next point with a trial probability $p(x)$ and replicate it with a weight given by (\ref{condmod}) divided by $p(x)$. Due to the shape of $U(x)$, the barrier to go to $x=+1$ is equal to $\beta/2$. With $T=0.02, \beta = 50$, this barrier is equal to 25. Thus starting with a population of points located in one side of the well, 
the probability to tunnel to the other side is very small, of the order of  $10^{-11}$. The partition function $Z_0$ is adjusted so that the population of points remains approximately constant.
In Fig. 1, we show the probability distribution obtained after $10^4$ Monte Carlo steps when $\lambda=0$ (no accelerating potential). 
\begin{figure}[htbp]
\centering
\includegraphics[width=2.2in]{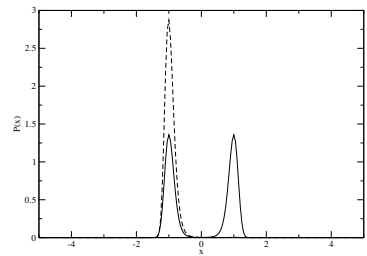}
\caption{Probability distribution with no acceleration factor after 10000 timesteps (dashed line) compared to the true Boltzmann distribution (solid line).}
\label{fig1}
\end{figure}
Obviously,  the system explores just one side of the well and cannot overcome the barrier with this number of MC steps.  

We have performed identical calculations with $\lambda=5$ with only 500 total MC steps. As can be seen on Fig. 2, the system is almost perfectly thermalized and the barrier has been easily overcome. Therefore, the sampling efficiency has been tremendously increased.
\begin{figure}[htbp]
\centering
\includegraphics[width=2.2in]{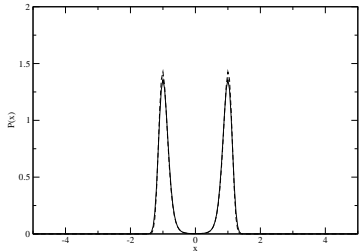}
\caption{Probability distribution after 500 timesteps (dashed line), with an acceleration factor $\lambda=5$ compared to the true Boltzmann distribution (solid line).}
\label{fig2}
\end{figure}

To assess more quantitatively the convergence of our method, we define the overlap of the final normalized probability distribution with the Boltzmann distribution
\be
\theta=\int dx P(x) e^{-\beta U(x)/2}/Z_1
\ee
with
\be
Z_1 = \int dx e^{-\beta U(x)/2}
\ee

This overlap $\theta$ is related to the distance $d$ between the final p.d.f. and the Boltzmann distribution through the identity
\be
d=2 (1-\theta)
\ee

An overlap of 1 means identical distributions, and an overlap close to 0 means very different distributions.
In Fig. 3, we plot the overlap as a function of the number of time steps, for $\lambda=0$ (solid curve) and for $\lambda=0.1$ (dashed curve). We use a small $\lambda$ otherwise the convergence is too fast to be seen on the scale of the figure. We see that the non-accelerated case is still far from convergence at 5000 steps, whereas the accelerated one has converged after about 1000 steps.
\begin{figure}[htbp]
\centering
\includegraphics[width=2.2in]{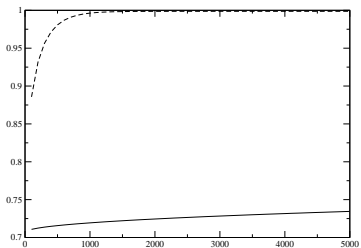}
\caption{Overlap of the probability distribution with the Boltzmann distribution as a function of the number of MC steps. The solid curve corresponds to $\lambda=0$ whereas the dashed one to $\lambda=0.1$.} 
\label{default}
\end{figure}

We have seen how by modifying in a simple way  the quantum Hamiltonian associated to the Brownian dynamics of a system, one can obtain a very large acceleration of the convergence of the probability distribution to the stationary distribution. The price to pay is that the new quantum Hamiltonian is no more local in space. However, it is still simple enough to be sampled efficiently.
We are presently trying to apply this method to more complex optimization problems and to generalize it to discrete variable problems.

\end{document}